# Embedded Deep Regularized Block HSIC Thermomics for Early Diagnosis of Breast Cancer

Bardia Yousefi, Hossein Memarzadeh Sharifipour, and Xavier P.V. Maldague

Department of Electrical and Computer Engineering, Laval University,
Quebec City (Quebec) G1V 0A6, Canada

*Abstract*— **Thermography has been used extensively as a complementary diagnostic tool in breast cancer detection. Among thermographic methods matrix factorization (MF) techniques show an unequivocal capability to detect thermal patterns corresponding to vasodilation in cancer cases. One of the biggest challenges in such techniques is selecting the best representation of the thermal basis. In this study, an embedding method is proposed to address this problem and Deep-semi-nonnegative matrix factorization (Deep-SemiNMF) for thermography is introduced, then tested for 208 breast cancer screening cases. First, we apply Deep-SemiNMF to infrared images to extract low-rank thermal representations for each case. Then, we embed low-rank bases to obtain one basis for each patient. After that, we extract 300 thermal imaging features, called thermomics, to decode imaging information for the automatic diagnostic model. We reduced the dimensionality of thermomics by spanning them onto Hilbert space using RBF kernel and select the three most efficient features using the block Hilbert Schmidt Independence Criterion Lasso (block HSIC Lasso). The preserved thermal heterogeneity successfully classified asymptomatic versus symptomatic patients applying a random forest model (cross-validated accuracy of 71.36% (69.42%-73.3%)).**

## I. Introduction

MATRIX factorization methods are known to be very effective in the detection of defects in the sequence of thermal images [1-6]. Such methods such as principal component analysis/thermography (PCA/PCT) [1], and NMF [6] have been used for the past two decades. The major challenge then remains how to deal with an abundance of low-rank approximations (bases) of high dimensional infrared imaging sequences accurately. This uses to enhance the visibility of the defects and impeding manual selection of the appropriate matrix (basis). Most dimensionality reduction and low-rank approximations methods suffer from a lack of final selection of the best representative factorized images.

PCT [1] decomposes the heat matrix, obtained by stacking the thermal sequence, into eigenvector and eigenvalue matrices. Alternative approaches tried to modify the PCT to improve its performance. Fixed eigenvector analysis method [7], incremental PCT [8], and candid covariance-free incremental principal component thermography (CCIPCT) [2] tried to overcome computational load by a fixed set of previously generated eigenvectors and covariance free approach, respectively. Sparse PCT [3,4] (or Sparse non-negative matrix factorization [9]) adds regularization terms to increase the sparsity in the analyses by restricting the solution domain and strive to convert this decomposition to a nonlinear method, which strengthened the important basis to detect better defective patterns. However, all these methods followed the fundamental principle of thermal non-destructive evaluation and were considered to be well behaved and slowly varying in time. Therefore, they select the initial eigenvector as the predominant temporal variations of the entire data set, whereas it cannot be usually generalized.

Non-negative matrix factorization (NMF) [10] limited the solution of such decomposition by additive constraints in basis and coefficient matrices that decompose an input matrix into a low-rank non-negative basis. Such constraints loosen for the basis in Semi-NMF, which is equivalent to a clustering problem [11]. NMF is used for the estimation of geometrical properties along with PCT and archetypal analysis (AA) [6]. An application of NMF in IRNDT applying two ways of computations using gradient descend (GD) [12] and non-negative least square (NNLS) [13] for evaluating cultural heritage objects and buildings illustrated the considerable performance of such algorithm for detecting subsurface defects [14]. Semi-NMF for thermography was very briefly discussed with an example of its application [15]. Deep-semiNMF [16] is also a modified version semiNMF algorithm by considering multiple hidden layers converting basis matrix to a set of basis matrices, which ultimately provides better clustering performance.

This work was supported by Canadian Tier-1 research chair in infrared and thermography.

B. Yousefi, H.M. Sharifipour and X.P.V. Maldague are with the Department of Electrical and Computer Engineering, Laval University. (e-mail: Bardia.Yousefi.1@ulaval.ca, Hossein.Memarzadehsharifipour.1@ulaval.ca, Xavier.Maldague@gel.ulaval.ca).

This is the authors version of the article published in IEEE Transactions on Instrumentation and Measurement, doi: 10.1109/TIM.2021.3085956.

Despite considerable developments in the methodology for detecting subsurface spatio-thermal patterns, automatically deep basis matrices have remained uninvestigated. Moreover, changing the bases in the deep-semiNMF method converted the problem to a segmentation problem, while selecting a predominant basis remained challenging. This paper uses low-rank deep-semiNMF for the first time in thermography for low-rank matrix approximation. Then, a membership embedding of the low-rank approximated matrices generates the predominant basis of matrix decomposition, which considers our major contribution.

Moreover, we proposed the application of thermal imaging features, called thermomics. Inspired by Radiomic features [17] in medical imaging. Then, we addressed the problem of the high dimensionality of thermomics by spanning them to Hilbert space and measuring the block Hilbert Schmidt Independence Criterion Lasso (HSIC Lasso) resulting in a significant reduction in the dimensionality of the features, by hundred times, which considers as our secondary contribution of this study. Besides, this study shows a confirmation of low-rank matrix factorization in thermal breast cancer screening and thermal heterogeneity detection for early diagnosis of breast cancer by identifying symptomatic patients.

In the next two sections, an overview of thermal transfer and the methodology of the approach using embedding for deep-semiNMF, thermomics, and block HISC lasso are described. The experimental and computational results, as well as the discussion, are then presented in Sections IV and V, respectively. Finally, the conclusions are presented in Section VI.

## II. THERMAL TRANSFER AND IMAGING SYSTEM

A thermal camera captures the spatial variation of temperature on the targeted region of interest (ROI) over time. This heat transient can be through active or passive thermography techniques. In general, the thermal transfer/heat conduction equation of a specimen can be summarized by the following equation:

$$\rho C_p \frac{\partial T}{\partial t} = k \frac{\partial^2 T}{\partial t^2} + \dot{q}$$

where $T = T(x, y, z)$ is a temperature field, $k$ is thermal conductivity constant from the material $(W/m.K)$. $\rho$ is the density $(kg/m^3)$, $C_p$ is specific heat $(J/kg.K)$, $\dot{q}(x,y,z,t)$ is the internal heat generation function per unit volume, in the passive thermography.

Applying infrared thermography on biological organs and tissues, as a complex structure, composed of fat, blood vessels, parenchymal tissues, and nerves with some uncertainty for the rate of blood perfusion and metabolic activity. Pennes' bioheat equation [18] provides accurate thermal computations and states as follows:

$$\rho_t c_t \left(\frac{\partial T_t}{\partial t}\right) = \nabla . (k_t \nabla T_t) + \omega_b c_b (T_a - T_t) + q_m$$

where $\omega_b$ represents the flow rate of blood, $q_m$ is the metabolic rate (heat generation), and $b$, and $a$ in $\omega_b c_b (T_a - T)$ the additive term stands for blood, and arteries (in targeted tissue), respectively.

## III. METHODOLOGY

### A. Sparse-NMF in thermography

NMF is comparable to the PCA decomposition yet the PCA's basis vectors can be negative. Input data $X$ is stacked vectorized thermal images, $\mathbf{X} \in \mathbb{R}^+_{MN \times \tau}$, and can be shown by a linear combination of $\tau$ bases, $\mathbf{B} = \{\beta_1, \beta_2, \ldots, \beta_p\}$, $\mathbf{B} \in \mathbb{R}^+_{MN \times \tau}$ and $\mathbf{A}$ coefficient matric, $\mathbf{A} \in \mathbb{R}^+_{\tau \times \tau}$, and $\mathbf{X} = \mathbf{BA}$ $s.t. \mathbf{A} \geq 0, \mathbf{B} \geq 0$. Many types of research have proposed ways and optimization solution such as GD algorithm and NNLS by Paatero and Tapper [19], multiplicative algorithm by Lee and Seung [10], a computational improved GD algorithm by Lin [12], and modified alternating non-negative least squares (ANNLS) by Liu et al. [13]. To compensate for the uniqueness of the decomposition and enforcing the representation of basis, sparseness constraints are proposed for NMF. A $\ell_1$ norm penalty term has been introduced by implementing the following equation:

$$C_{\text{Sparse NMF}} = \frac{1}{2} \|\mathbf{X} - \mathbf{BA}\|_F^2 + \lambda \|\mathbf{B}\|_1$$

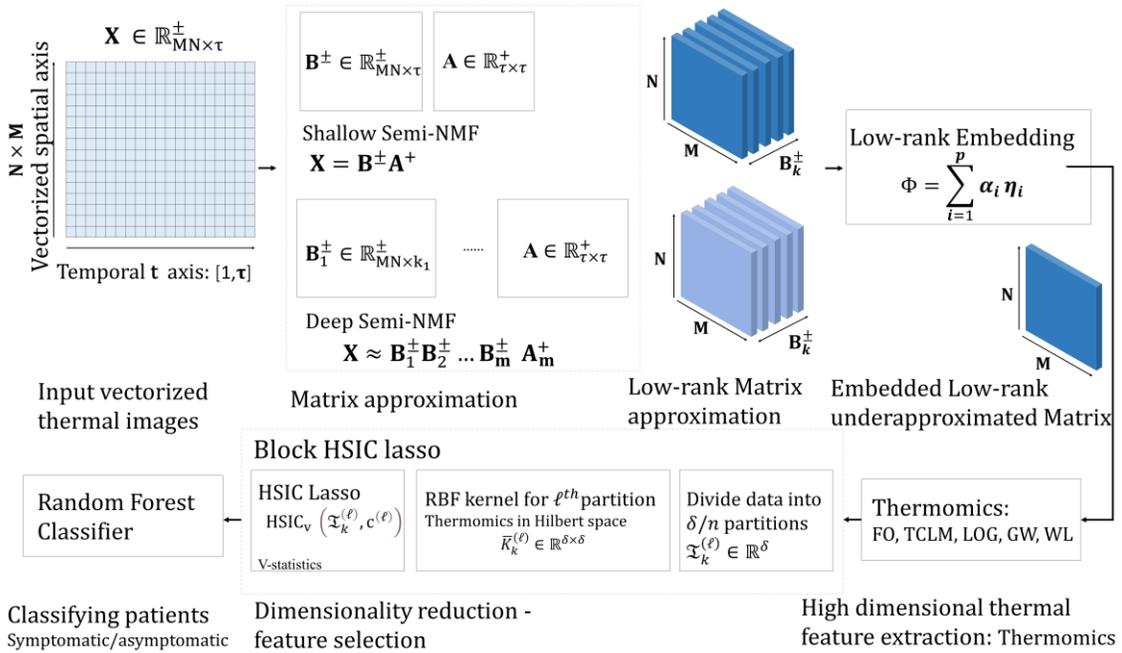

**Fig. 1.** Workflow of the proposed approach using Deep-SemiNMF versus shallows SemiNMF, with low rank embedding method, and block HSIC lasso are presented.

Let $\|\mathbf{B}\|_p$ is the $\ell_p$-norm of $\mathbf{B}$ given by $\ell_p = \sum_{d,m} \|\mathbf{B}_{d,m}\|_p^{1/p}$ similar to usual $\ell_1$ penalty term to imitate the $\ell_0$ behavior [20] to calculate $\mathbf{B}$ for convex $\mathbf{A}$. It is the unconstrained least squares minimization with $\ell_1$-norm constraint is also referred to as the least absolute shrinkage and selection operator (LASSO) [28]. Like Sparse PCT, low-rank sparse NMF is obtained by selecting $k$ bases that correspond to the highest coefficients.

*B. Deep Semi-NMF in thermography*

Semi non-negative matrix factorization (Semi-NMF) [11] relaxes the constraints of NMF for the basis matrix while keeping the non-negative restriction for the coefficient matrix.

$$\mathbf{X}^\pm \approx \mathbf{B}^\pm \mathbf{A}^+$$

This is also can be perceived as clustering with sets of grouping centroids, $\mathbf{B} = [B_1, B_2, \ldots, B_k]$, and indicators, $\mathbf{A} = [A_1, A_2, \ldots, A_n]$. Deep Semi-NMF model expands basis factors to $m+1$ factors, as it is shown as follows:

$$\mathbf{X}^\pm \approx \mathbf{B}_1^\pm \mathbf{B}_2^\pm \ldots \mathbf{B}_m^\pm \mathbf{A}_m^\pm$$

This allows for a hierarchy of bases ($m$ layers) using a chain of factorization the basis matrices to have pyramid representations of thermal sequences (see Fig1), which can be presented as follows:

$$\mathbf{A}_{m-1}^+ \approx \mathbf{B}_m^\pm \mathbf{A}_m^+$$
$$\vdots$$
$$\mathbf{A}_2^+ \approx \mathbf{B}_3^\pm \ldots \mathbf{B}_m^\pm \mathbf{A}_m^+$$
$$\mathbf{A}_1^+ \approx \mathbf{B}_2^\pm \ldots \mathbf{B}_m^\pm \mathbf{A}_m^+$$

which involves non-negative constraints for coefficients for all the layers, $(\mathbf{A}_1^+, \ldots, \mathbf{A}_{m-1}^+)$.

A Deep Semi-NMF model decomposes the heat map made a low-dimensional representation of thermal sequence with their dominating characteristics. This provides complex hierarchical information obtained from the data corresponding to implicit lower-level hidden attributes, where here $\mathbf{B_1B_2B_3}$ is corresponding to the mapping of thermal heterogeneity related to breast cancer. Fig. 2 shows low-rank approximation of thermal sequence determined using sparseNMF and DeepSemiNMF matrix factorization techniques for symptomatic and healthy cases.

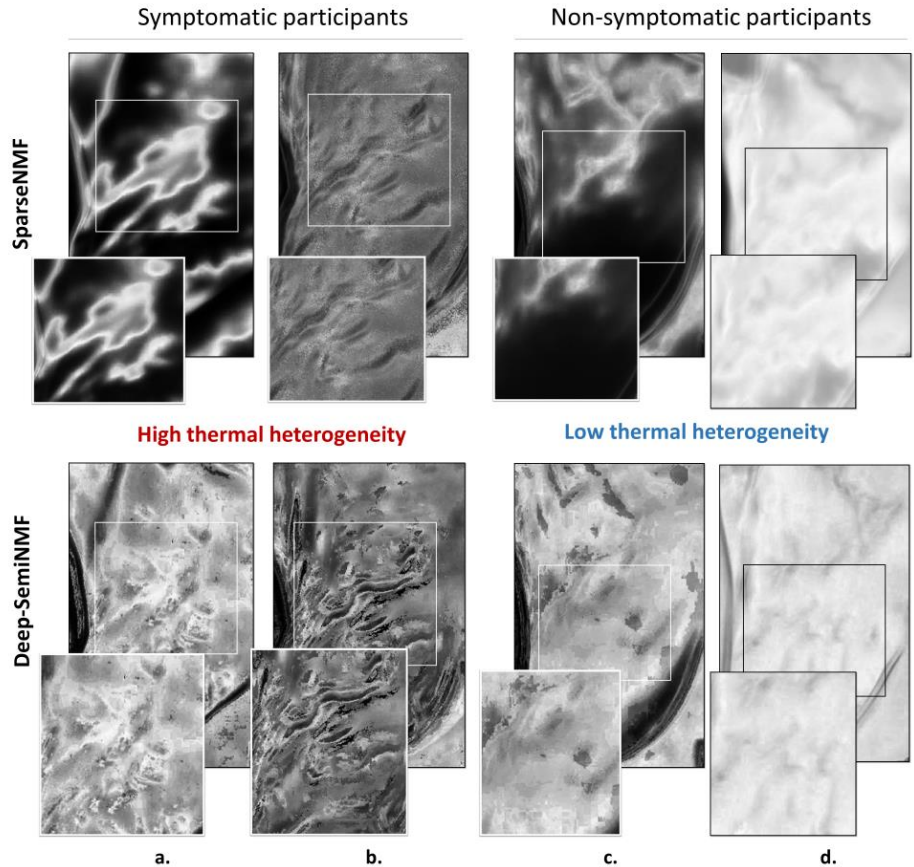

**Fig. 2.** Low-rank approximation of thermal sequence determined using sparseNMF and DeepSemiNMF matrix factorization techniques. Columns (**a-b**) show symptomatic patients (diagnosed by mammography as cancer patients or healthy with symptoms), whereas columns (**c-d**) show the result of methods for healthy participants.

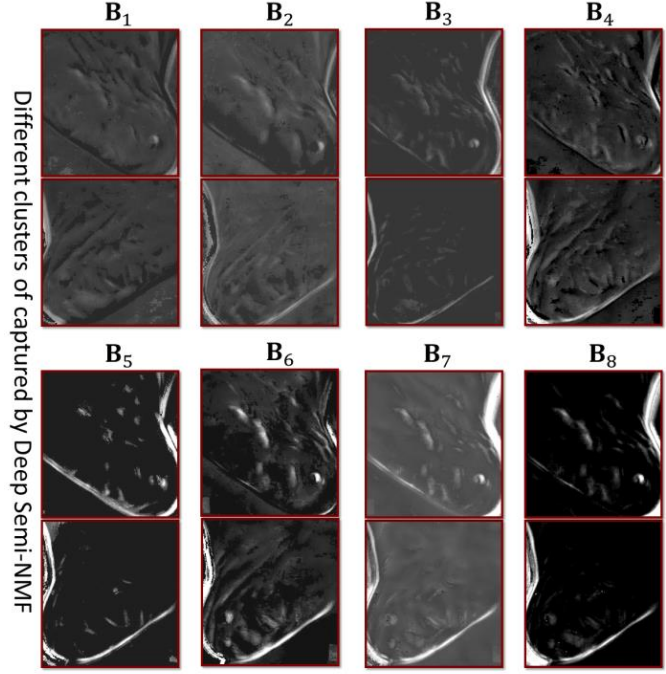

**Fig. 3.** Different clusters of thermal heterogeneity captured by deep SemiNMF for a breast cancer patient.

$$C_{deep} = \frac{1}{2}\|X - B_1 B_2 \ldots B_m A_m\|_F^2$$
$$= tr[X^T X - 2X^T B_1 B_2 \ldots B_m A_m + A_m^T B_m^T B_{m-1}^T \ldots B_1^T B_1 B_2 \ldots B_m A_m]$$

where $C_{deep}$ is the cost function used as the reconstruction error used for the pretraining and fine-tuning the weight for the layers. We pre-train every layer for the approximation of bases and coefficients and expedite training time. We break down the overall decomposition by decomposing $X \approx B_1 A_1$ then decompose $B_1$, until all the layers get trained. Fig. 3 shows different clusters of thermal heterogeneity captured by deep SemiNMF for a breast cancer patient.

### C. Low-rank embedding

The low-rank matrix approximation using NMF was represented by $k$ non-negative basis and showed by $k$ predominant segments of the vectorized thermal images. A set of low-rank representation of basis vector is showed $\mathbf{B} = \{\beta_1, \beta_2, \ldots, \beta_p\}$, where $\mathbf{B} \in \mathbb{R}^{NM \times \tau}$. SemiNMF like NMF shows slight sparsity among its bases [11], which is enhanced using Deep-semiNMF by multiple decompositions of the bases [22]. Here, a modified sparsity measure motivated by [23] along with the definition of embedding on the low-rank approximation is presented.

**Definition 1**. Sparsity, $\xi(.) \in [0,1]$, defines as the proportion of its zero entries on the dimensionality of the matrix, $Q \in \mathbb{R}^{r \times c}$, and shows as follow:

$$\xi(Q) = \frac{1}{rc}\sum_{i=1}^{r}\sum_{j=1}^{c}\gamma_{i,j}, \quad \gamma_{i,j} = \begin{cases} 1 & f_{i,j} = 0 \\ 0 & f_{i,j} \neq 0 \end{cases}$$

**Theorem**. For any non-negative low-rank ($\ell$) approximation $\alpha \in \mathbb{R}_{m \times k}^{+}$, $\beta \in \mathbb{R}_{k \times n}^{+}$ of $X \in \mathbb{R}_{m \times n}^{+}$, we have:
$$\xi(\alpha) + \xi(\beta) \geq \xi(X) \text{ such that } 1 \leq \ell \leq k.$$

**Proof.** Let consider rank-one approximation of input data, $X$, the number of nonzero elements in the approximated matrix, $\alpha^1 \beta^1$, are equal to multiplication of non-zero elements in two vectors $\alpha^1$ and $\beta^1$. Hence from Definition 1, we have:
$$1 - \xi(\alpha^1 \beta^1) = (1 - \xi(\alpha^1))(1 - \xi(\beta^1))$$
which implies $\xi(\alpha^1 \beta^1) = \xi(\alpha^1) + \xi(\beta^1) - \xi(\alpha^1)\xi(\beta^1) \leq \xi(\alpha^1) + \xi(\beta^1)$. Moreover, since this is an under-approximation inherently satisfies $0 \leq \alpha^1 \beta^1 \leq X$, then $\xi(\alpha^1) + \xi(\beta^1) \leq \xi(\alpha^1 \beta^1) \leq \xi(X)$. Considering the recursive implementation of the NMF, $\hat{X} = X - \alpha^1 \beta^1$, and repeat the previous procedure for $\alpha^1 \beta^1$ **of** $\hat{X}$, which ultimately gives $\xi(\alpha^1) + \xi(\beta^1) \geq \xi(\hat{X})$, till $\alpha^1 \beta^1 \nleq \hat{X}$.

This proved the claim, and now here we show the effects of two additive constraints $1 \leq \ell \leq k$, and $\alpha \neq 0$ suppose $\ell \neq k (1 \leq \ell < k)$, approximation might not be equal to the input data $\xi(\alpha) + \xi(\beta) \neq \xi(\alpha\beta) \neq \xi(X)$, where $\alpha$ and $\beta$ are constructive vectors of input matrix $X$, which against the initial assumption. □

Having the sparsity among the non-negative basis implies high segmentation characteristics particularly in the imaging-based data (see [10]). Here, we argue that the final response of the low-rank representation is an aggregation of different reconstructing bases. We define the membership function to highlight such property to aggravate and integrate the overall representation of under-approximation in the bases.

**Definition 2**. The low-rank embedding, $\Phi$, defines by aggregating membership calculated for $p$ bases of $X$, $\mu_p$, multiply by basis itself, $\beta_i$, and defined as:

$$\Phi = \sum_{i=1}^{p} \alpha_i \, \eta_i$$

where $\eta_i$ is a membership of basis $\beta_i$ and defines by:

$$\eta_i = e^{\frac{\beta_i - \mu}{\sigma}}$$

Let $\mu$, $\sigma$ mean (average) of thermal basis and standard deviation of $i^{th}$ −basis in the calculation. The intuition behind using this embedding concerns highlighting the thermal difference in exponential order, which enhances the thermal heterogeneity in the aggregated representative image, called the *avatar*.

### D. Thermomics: Thermal Imaging features

Upon creating low-rank embedded representative of thermal images in the breast screening, a set of imaging features involving statistical analyses, texture and grey level intensity, and spatiotemporal features from the ROI.

Statistical analyses were performed to measure the heterogeneity of the breast area using texture analyses, ten features. Having thermal measure encoded in the grey level intensity, the grey (thermal) level co-occurrence matrices (TLCMs) of the breast area as our ROI was calculated. For each patch, TLCM with a horizontal offset of 4 (two distances [0,5] and five angles [0, $\pi$]) is computed. Using the TLCM information, many analyses were conducted to measure the level of contrast, dissimilarity, correlation, energy, and homogeneity among the pixels in the ROI.

Some spatiotemporal features such as Gabor wavelet, Laplacian of Gaussian, and wavelet were also measured for the targeted area. Overall, 300 features were extracted from thermal imaging to assess the thermal heterogeneity of the breast area.

### E. Hilbert-based dimensionality reduction

Having 300 features for each embedded thermal image enforces using dimensionality reduction to provide a solution to the *curse of dimensionality*. For that, we select features with higher variance and estimated their dependency by spanning them onto Hilbert space and measuring their distributions using the least absolute shrinkage and selection operator (lasso) in the Block Hilbert-Schmidt Independence Criterion (Block HSIC-lasso) [24,25]. This algorithm first partition the training set to $\delta/n$ partitions $\left\{\{x_i^\ell, c_i^\ell\}_{i=1}^{\delta}\right\}_{\ell=1}^{n/\delta}$, where $\delta$ samples stay in every block, $\delta \ll n$. $x_i^\ell$ and $c_i^\ell$ are $\ell^{th}$ attribute, in this paper $\ell^{th}$ thermomic batch, and its clinical scoring. Here $\delta = 20$ and $n = 300$. Then having V-statistics, the block HSIC can be estimated as follows:

$$\text{HSIC}_b(\mathfrak{T}_k, c) = \frac{\delta}{n} \sum_{\ell=1}^{n/\delta} \text{HSIC}_v\left(\mathfrak{T}_k^{(\ell)}, c^{(\ell)}\right)$$

where $\mathfrak{T}_k^{(\ell)} \in \mathbb{R}^\delta$ symbolizes the $k^{th}$ thermomic vector of the $\ell^{th}$ partition (the data is spanned on to Hilbert space using Radial Based Function Gaussian kernel for pairs of thermomics $K: x_i^{(k)}, x_j^{(k)} \mapsto \exp\left(-\frac{\|x_i^{(k)} - x_j^{(k)}\|_2^2}{2\sigma^2}\right)$). If $\overline{K}_k^{(\ell)} \in \mathbb{R}^{\delta \times \delta}$, $K: \mathbb{R}^d \times \mathbb{R}^d \rightarrow \mathbb{R}$ represents the restriction of $\overline{K}_k^{(\ell)}$ to the $\ell^{th}$ partition, and by $\overline{C}^{(\ell)} \in \mathbb{R}^{\delta \times \delta}$, $C: c \times c \rightarrow \mathbb{R}$ the restriction of $C$ to the $\ell^{th}$ partition. Here, for the given Gram matrices corresponding to $k^{th}$ thermomics, $K_k$, and output $C$, we will have:

$$\text{HSIC}_v\left(\mathfrak{T}_k^{(\ell)}, c^{(\ell)}\right) = tr\left(\overline{K}_k^{(\ell)} \overline{C}^{(\ell)}\right) = vec\left(\overline{K}_k^{(\ell)}\right)^T vec\left(\overline{C}^{(\ell)}\right).$$

where $\overline{K} = AKA / \|AKA\|_F$. The Block HSIC lasso obtains by following optimization problem:

$$\max_{w \geq 0} \sum_{k=1}^{d} w_k \text{HSIC}_b(\mathfrak{T}_k, c) - \frac{1}{2} \sum_{k'=1}^{d} w_k w_{k'} \text{HSIC}_b(\mathfrak{T}_k, \mathfrak{T}_{k'}) - \lambda \|w\|_1$$

where $w_k$ is the weight of $k^{th}$ thermomics. To apply Block HSIC-lasso, we create a scoring technique to weight each patient in the breast cancer screening by clinical information (*i.e.*, symptoms, physical changes in the shape of the breast). Having such information, we reduce the dimensionality and use a binary classification random forest model to classify the patients based on whether they are symptomatic or healthy patients. Having thermomics ranked, we stratified the participants based on the highest selected thermomic and compared it based on the ground truth data. Man-Whitney U test was performed to determine the statistical significance

TABLE I
CLINICAL INFORMATION AND DEMOGRAPHICS OF THE BREAST CANCER SCREENING DATABASE USING THERMAL IMAGING.

| DMR - Database for Mastology Research | | |
|---|---|---|
| **Age** | Median (±IQR) | 60 (25,120) |
| **Race** | Caucasian | 77 (37%) |
| | African | 57 (27.4%) |
| | Pardo | 72 (34.6%) |
| | Mulatto | 1 (0.5%) |
| | Indigenous | 1 (0.5%) |
| **Diagnosis**[1] | **Healthy(non-symptomatic)**[2] | 128 (61.5%) |
| | | 80 (38.5%) |
| | **Symptomatic Sick**[3] | 36 (17.3%) |
| **Family history** | Diabetes | 52 (25%) |
| | Hypertensive | 5 (2.4%) |
| | Leukemia | 1 (0.5%) |
| | None | 150 (72.1%) |
| **Hormone therapy (HT)** | Hormone replacement | 38 (18.3%) |
| | None | 170 (81.7%) |

[1] *This diagnosis performed with mammography as ground truth in this Dataset.*
[2] *Healthy term is used as non-cancerous and non-symptomatic patients.*
[3] *We use the term "sick", which includes different types of breast cancer patients diagnosed by mammographic imaging.*

between two groups' discriminations (symptomatic versus healthy participants).

## IV. RESULTS

The proposed method for thermal pattern detection was examined by thermal breast cancer screening datasets. The results of the low-rank approximation using Deep-semiNMF were then compared to other state-of-the-art thermal low-rank matrix approximation algorithms. The results were assessed in the presence of additive Gaussian noise to determine the stability of methods facing noise by calculating the signal-to-noise ratio (SNR).

### A. Patient population and Infrared breast cancer study data

208 participants who were healthy (without symptoms) or sick (diagnosed by mammographic imaging as breast cancer cases or non-cancerous but with symptoms) were employed for breast screening. The median age in our study sample was 60 years, and the participants comprised 77 (37%) Caucasian, 57 African (27.4%), 72 Pardo (34.6%), 1 Mulatto (0.5%), and 1 indigenous (0.5%) women. Among the participants, 52 had a history of diabetes in their families (25%), and 38 were undergoing hormone replacement (18.3%). All patients had infrared images obtained by the following acquisition protocol: images have a spatial resolution of 640×480 pixels and were captured by a FLIR thermal camera (model SC620) with a sensitivity of less than 0.04ºC range and capture standard of −40ºC to 500ºC [26]. Table 1 shows the clinical information and demography of the cohort.

### B. Results of low-rank Deep-semiNMF

Eight low-rank matrices were extracted from the 23 initial thermal sequences by using Deep-semiNMF. Some representative results of the low-rank approximation manually selected for our study cohort are shown in Fig.3. Low-rank approximation in the sequence of thermal images resulted in a heterogeneous breast area for more than 80 participants for breast cancer screening (sick and healthy with symptoms or abnormality, Fig.2.a-b). Thermal patterns showed more homogeneous structures were detected among the healthy participants (healthy with no symptoms, Table 1, Fig.2.c-d).

TABLE II
THE RESULTS OF BLOCK-HSIC LASSO TO SELECTING THERMOMIC FEATURES USING THEIR RELATEDNESS SCORING.

| Block-HSIC Lasso scoring of the thermal features | | |
|---|---|---|
| Methods | Thermomics | Relatedness Score |
| CCIPCT | First order-interqu10 | 1.00 |
|  | Dissimilarity 1-π | 0.91 |
|  | First order-min | 0.87 |
| PCT | Contrast 2-0 | 1.00 |
|  | Contrast 1-3π/4 | 0.58 |
|  | First order -Gray range | 0.51 |
| NMF | LOG-1-0-firstorder-interqu10 | 1.00 |
|  | ASM 1-π | 0.84 |
|  | Dissimilarity 5-3π/4 | 0.64 |
| Sparse PCT | First order-MAD | 1.00 |
|  | Contrast 1-3π/4 | 0.61 |
|  | Dissimilarity 4-π | 0.54 |
| Sparse NMF | ASM 5-3π/4 | 1.00 |
|  | Correlation 1-π/2 | 0.53 |
|  | Energy 5-π/2 | 0.47 |
| Deep-SemiNMF | Dissimilarity 2-π | 1 |
|  | Correlation 3-π/4 | 0.61 |
|  | Homogeneity 4-π/2 | 0.54 |

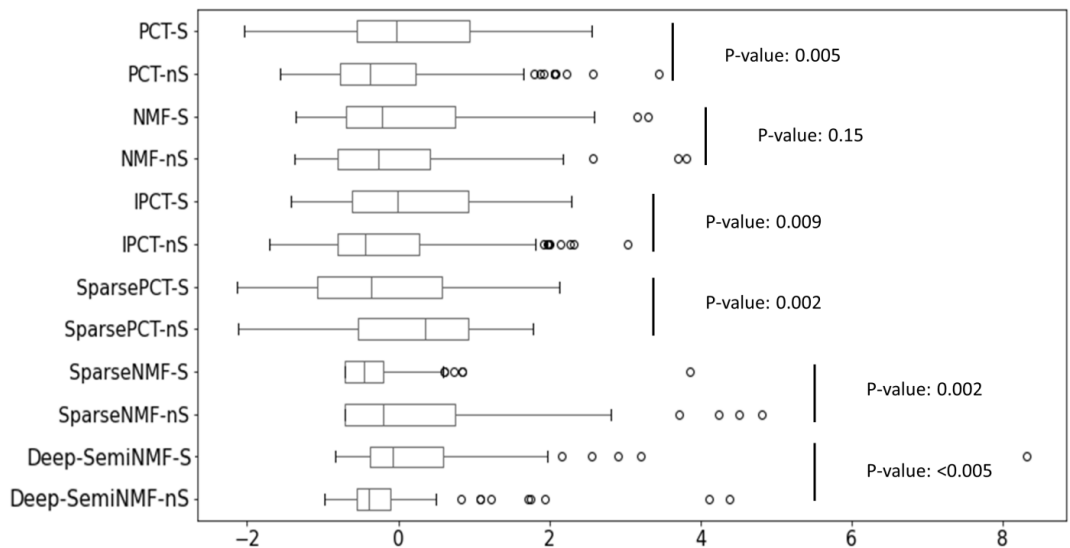

**Fig. 4.** Stratification of participants into symptomatic (abnormal-"S" suffix) and non-symptomatic (healthy-"nS" suffix) groups for each matrix factorization algorithm as presented by boxplots and Mann–Whitney U test.

## C. Results of low-rank embedding

Eight low-rank approximated matrices obtained by matrix factorization approaches were embedded using the proposed approach. To determine the level of thermal heterogeneity in the breast area, we used the reference label, attached between the breasts, to normalize the representation for each image. Thermal heterogeneity significantly increased after embedding, which is much more significant for symptomatic patients than healthy participants (see Fig.2).

## D. Thermomic features

Three hundred thermomic features have been extracted from thermal imaging from the targeted ROI (solely breasts area) in four categories: first-order statistics, texture, intensity, spatiotemporal features. Statistical features were involved in calculating 10 and 90 percentage of percentile, maximum, minimum, median, mean, interquartile range, gray level range, mean absolute deviation (MAD), standard deviation, and skewness of intensity within ROI. Many analyses were conducted with the TLCM information to measure the level of contrast, dissimilarity, correlation, energy, and homogeneity among the pixels in the ROI. The LOG, Gabor wavelet, and wavelet were computed and were stacked in a matrix with 300 thermomics correspondings to each thermal image. To alleviate such collinearity among the thermomics, Block HISC Lasso using the Gaussian kernel reduced the dimensionality of data 100 times by selecting three of the most effective thermomics using relatedness scoring (Table 2).

We stratified the participants based on the highest relatedness scoring properties and compared them with the ground truth data based on mammography information. Mann–Whitney U test was performed to determine the statistical significance of stratification between two groups (symptomatic versus healthy participants) using targeted thermomics (see Fig.4).

Deep-semiNMF showed statistical significance in separating the two groups of participants when splitting on the targeted thermomics (p < 0.005, Fig.4). This also compared to Sparse-NMF, Sparse- PCT, PCT, and CCIPCT and showed statistical significance in discriminating symptomatic and healthy participates. NMF did not show strong stratification ability (p = 0.15, Fig.4). Deep-semiNMF showed slightly higher separation strength compared to other approaches (Fig.4).

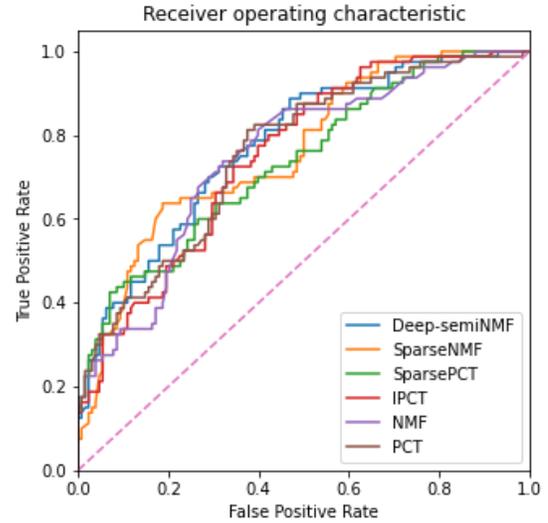

**Fig. 5.** ROC graph for different matrix factorization approaches for multivariate covariate random forest (binary classification).

To examine the hypothesis that the thermal heterogeneity can be used as a biomarker to stratify among participant, a random forest model fitted for leave one out cross-validation multivariate covariates, the best three selected thermomics and resulted in 71.36% (69.42%-73.3%) accuracy for Deep-semiNMF and 72.33% (69.42%-74.27%) for sparse PCT, which were challenged by other matrix approximation techniques such as Sparse-NMF, PCT, and CCIPCT. After Deep NMF and Sparse PCT, the highest accuracy was belonging to PCT, with values of 70.63% (67.48%-74.27%). CCIPCT and NMF were both showed an accuracy of about 69.42% (67.48%-71.36%) and 66.5% (63.59%-69.9%), respectively (Table 3). The ROC graph of comparative analyses is shown in Fig.5.

## E. Robustness in thermal defect patterns

Noise in the sequence of thermal images is an unequivocal problem, which depends on the acquisitions' condition. It may aggravate having *vivo* specimens as subjects of the study. The robustness of the deep-semiNMF in comparison with other discussed algorithms versus additive Gaussian noise input and measured using SNR by the following equation (from [27]):

TABLE III
THE RESULTS OF RANDOM FOREST CLASSIFICATION FOR THE BASELINE AND LEAVE-ONE-OUT CROSS-VALIDATION MODEL

| | Multivariate full covariates binary classification using random forest classifier | | | |
|---|---|---|---|---|
| Methods | Without Embedding (%) | t-test t-statistic, two-tailed p-value | With Embedding (%) | t-test t-statistic, two-tailed *p*-value |
| CCIPCT | 65.1 (62.1-66.9) | 7.9, <0.0005 | 69.4 (67.5-71.4) | 2.8, 0.005 |
| PCT | 65.3 (63.1-67.9) | 6.8, <0.0005 | 70.6 (67.5-74.3) | 1.1, 0.29 |
| NMF | 64.6 (63.6-65.5) | 7.5, <0.0005 | 66.5 (63.6-69.9) | 7.2, <0.0005 |
| Sparse PCT | 67.5 (65.1-69.4) | 2.9, 0.003 | 72.3 (69.4-74.3) | 0.9, 0.35 |
| Sparse NMF | 62.6 (62.2-66.51) | 10.7, <0.0005 | 64.6 (61.-68.9) | 10.3, <0.0005 |
| Deep SemiNMF | 69.9 (66.5-71.4) | - | 71.4 (69.4-73.3) | - |

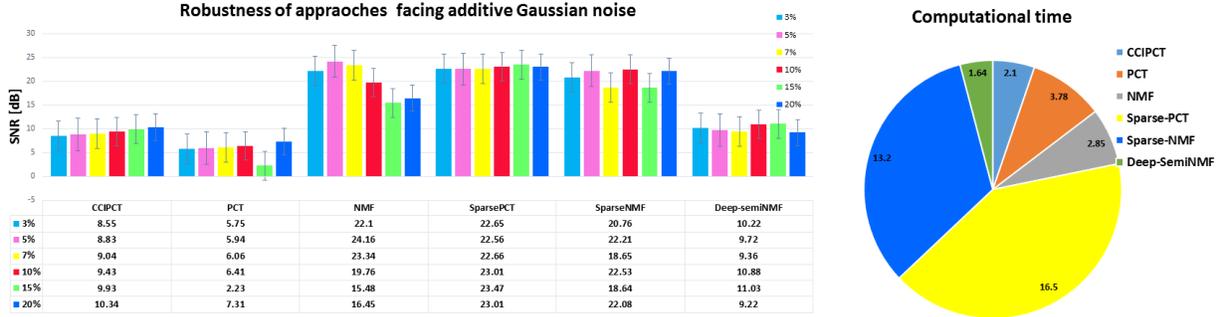

**Fig. 6.** SNR analysis (left side) and computational time (right side) of the state-of-the-art approaches in thermography.

$$\text{SNR} = 10\log_{10}\frac{|\mu_S - \mu_N|^2}{\sigma_N^2}$$

where $\mu_S$ and $\mu_N$ are the average levels of signal in ROI and noise, respectively. $\sigma_N$ is the standard deviation of noise in the reference or sound region of the image. Many other definitions have been established for the SNR. In this study, we follow one definition for the entire comparative analysis (following [27]). In general, low-rank matrix factorization prevents additive noise in the spatio-thermal defect pattern detection due to low-rank noise reduction in such algorithms.

The robustness of each method was determined by the SNR metric when Gaussian noise was added to the input image set. Low-rank matrix approximation was performed for each case study while the noise level was increased from 3% to 20%. The results indicate that Sparse-PCT and sparse-NMF were had the highest robustness compared to other approaches because of the additive regularization parameters. This relatively induces the nonlinear behavior of these techniques. CCIPCT and Deep-semiNMF methods were more stable than SNR of input with higher additive noise; however, these methods had comparatively lower overall SNR than sparse-based methods. Fig. 6 shows the robustness of the proposed method using SNR versus state-of-the-art approaches with computational time.

The computational process was partially performed with a PC (Intel Core 2Quad CPU, Q6600, 2.40 GHz, RAM 8.00 GB, 342-bit Operating System) by using Python programming language.

## V. Discussion

In this study, we used embedded deep-semiNMF to extract thermal patterns for infrared diagnostic systems for thermography imaging. This study was designed based on the general trend of dimensionality reduction and defect detection methods (such as [1-4]) but solved the problem of multiple low-rank approximations. This study showed a great possibility to identify potential patients with breast cancer using non-invasive, faster, and more cost-efficient imaging modalities.

Embedding analysis not only added to Deep-semiNMF but also added to sparse NMF, sparse PCT, NMF, CCIPCT, and PCT and showed significant improvements in stratifying symptomatic patients from healthy participants (Fig.2, Fig.3, and Table 3) similarly to other approaches but NMF (Fig.3). Moreover, Deep-semiNMF showed higher accuracy than other approaches in finding heterogeneous thermal patterns, which might be due to the nature of hidden basis matrices and recursive training of the network. Sparse PCT and PCT also showed better accuracies than NMF and Sparse-NMF, which indicates non-negative constraints were not in the favor of detecting symptomatic cases.

The application of thermomics increased the dimensionality of the input thermal imaging and intensify the possibility of overfitting the random forest model, *curse of dimensionality* [28-34]. The Block-HSIC lasso reduced the dimensionality by removing the redundancy among the features by spanning thermomics to higher dimensional space using RBF Gaussian kernel and measuring HSIC lasso, which increases the robustness of feature selection versus outliers.

Thermal and infrared imagery has been used to determine breast abnormality for the past several years [35]. Discussions about the better positions for such imaging acquisitions [36] and about the reliability of this modality for detecting the modality [37] have been reported. However, the association of low-rank approximation of thermal heterogeneity with breast abnormality has not been discussed in literature nor embedding analyses of the low-rank matric approximation, which increases the novel of this contribution to the field.

One limitation for applying the presented models is related to data, despite a considerable number of cases. Having a baggier cohort of patients increases the statistical power of such analysis by increasing the possibility of independently validating the system (substitute of cross-validation). The other limitation may be using limited thermomic features. Having more thermomics helps to test the strength of the Block-HSIC lasso approach to select better features that lead to capturing better thermal characteristics.

The presented techniques offer some advantages. First, embedded Deep-semiNMF not only performed a low-rank approximation of input data but also eliminates the manual selection of the *avatar*. This considers as a significant help to such analyses and frequently mentioned to be a problem in the automated system [1-4,6,14,15]. Second, the proposed method considerably alleviates the effect of motion artifacts and noise, which can be substantial improvements in infrared thermography applications. To the best of our knowledge, this is the first time such a study has been performed introducing Deep-SemiNMF, low-rank embedding, and block HSIC lasso for Thermomics.

## VI. Conclusion

This study addressed one of the biggest challenges in the low-rank matrix approximation, which is selecting the best representative matrix approximation of the input data by proposing a low-rank embedding approach. We tested our method using a Deep-SemiNMF method tested for 208 thermal breast cancer screening cases. Moreover, we extract 300 thermal imaging features to encode imaging information for the automatic diagnostic model. We reduced the dimensionality of features by spanning thermomics to Hilbert space using RBF kernel and applying block HSIC Lasso. Deep-SemiNMF showed significant statistical power, using MWU-test, to classify symptomatic from asymptomatic participants ($p$-value $< 0.005$). We assessed the appropriateness of these approaches compared to state-of-the-art thermographic methods, such as PCT, CCIPCT, NMF, Sparse PCT, and Sparse NMF. The results indicated that embedded Deep-SemiNMF and embedded Sparse PCT have significant performance in preserving thermal heterogeneity and discriminating between symptomatic and healthy participants yielded the accuracies of 71.36% (69.42%-73.3%), and 72.33 (69.42-74.27), respectively. Deep-SemiNMF also performed with the lowest computational time than other methods.

Future works should involve more thermomics extracted from the low-rank approximation to increase the potential of assessing the entire thermal characteristics of cancerous parenchymal tissues. Moreover, a larger infrared imaging cohort will help to validate the performance of this system and can further confirm the strengths and limitations of this approach.